\documentclass[12pt]{article}
  
\usepackage[a4paper, total={7in, 9.5in}]{geometry}
\usepackage[super,square]{natbib}

\usepackage{doc}
\usepackage{subcaption}
\usepackage[version=3]{mhchem}
\usepackage{bm}
\usepackage{url}
\usepackage[version=3]{mhchem}
\usepackage{tikz}

\usepackage{sectsty}
\usepackage{etoolbox}
\sectionfont{\fontsize{16}{15}\selectfont}
\subsectionfont{\fontsize{16}{15}\selectfont}
\allsectionsfont{\mdseries}
\usepackage{graphicx}
\usepackage{epstopdf}
\title{\Large An Investigation into the Kinetics of $Li^+$ Ion Migration in Garnet-Type Solid State Electrolyte: $Li_7La_3Zr_2O_{12}$ }
\author{Aditya Muralidharan}
\date{}

\begin{document}

\maketitle

\hrule
\abstract{An all solid-state thin film lithium ion battery has been touted the holy grail for energy storage technology ever since the inception of the first one in $1986$ by Keiichi Kanehori. Solid-state batteries provide the distinct advantage of outperforming current technology by having a simpler composition, being easier and cheaper to manufacture, safer and having a higher theoretical gravimetric and volumetric energy density.
The commercialization of this technology however, is plagued by its own set of challenges, primarily low ionic conductivity and interfacial stability of the solid-state electrolyte separating the anode and cathode, a small electrochemical window and sub-par mechanical properties. In the last decade considerable progress has been made in remedying these issues with garnet-type electrolytes, especially $Li_7La_3Zr_2O_{12}$ $(LLZO)$, having emerged the leading contender. This has prompted renewed effects into the field of solid-state ionic's and  maximizing the ionic conductivity of $LLZO$ by modifying its properties, primarily by means of doping with a varying degree of success. Carving a clear road ahead requires an in-depth understanding of the origin of the high $Li^+$ ion conductivity, the primary means of investigating which is by first-principle methods. In this term paper we try to gain insight into the origin of mechanisms at play that drive the collective migration of $Li^+$ ions in $LLZO$ using a first-principles approach, to gain a deeper understanding and appreciation for optimizing its properties for use in next-generation energy storage systems. 
}
\\
\hrule
\section*{\normalfont Introduction }
\label{introduction}

Solid-state lithium-ion batteries are expected to overtake conventional ones sometime within the next decade, primarily due to the extensive research being done to overcome safety limitations with conventional liquid electrolyte based cells.\cite{wang}\cite{murugan} These can be broadly classified into organic; usually based on polymer electrolytes, or inorganic like those based on ceramic materials.\cite{wang} Among ceramics, there has been particular interest in "fast lithium-ion conductors" which are materials that while being electrical insulators can effectively conduct ions.\cite{wang}\cite{murugan} Garnets are a specific class of ceramic orthosilicates having the general molecular formula $A_3B_2(SiO_4)_3$, where $A$ is a 8-fold coordinated site and $B$ is a 6-fold coordinated site. When the silicon atoms in such a structure are replaced by lithium atoms the resulting structure is termed a lithium garnet.\cite{murugan} Furthermore, it is possible to increase the lithium content per unit to more than 3 by manipulating the valencies of the $A$ and $B$ atoms.\cite{murugan} These "stuffed lithium garnets" $(Li^+>3\rightarrow7)$ are usually exponentially faster lithium-ion conductors and maximally stuffed ones have the general formula $Li_7A_3^{III}B_2^{IV}O_{12}$.\cite{wang}\cite{murugan}\cite{cussen}

Among these $Li_7La_3Zr_2O_{12}\;(LLZO)$ has emerged as the most promising candidate.\cite{wang}\cite{murugan} Found to exist in two different polymorphs, cubic $(c-LLZO)$ and tetragonal $(t-LLZO)$, the cubic phase exhibits extremely high ionic conductivity $(10^{-4}S-10^{-3}S)$, and stability against a lithium metal anode compared to the tetragonal phase  $(10^{-8}S-10^{-6}S)$ , while being extremely cost effective to manufacture.\cite{wang}\cite{murugan}\cite{cussen} Studies have also shown that the desired high conductivity $c-LLZO$ is stabilized by the presence of trace amounts of aluminium ions, accidentally introduced by diffusion from the alumina crucibles used for sintering.\cite{murugan}\cite{xie} This has warranted great interest in optimizing the stability and conductivity of $c-LLZO$, primarily by varying its dopant composition.\cite{wang}\cite{murugan} Consequently, it is imperative that the underlying mechanism of lithium ion migration within the lattice be unravelled and understood, this is where first-principle methods like molecular dynamics $(MD)$ and nudged elastic band $(NEB)$ light a path forward.\cite{meier}\cite{he} By understanding interactions of the mobile lithium ions with the rest of the lattice, it becomes easier to pinpoint the reason for migration and therefore modify it to our benefit.\cite{meier} These methods also help us make theoretical predictions about modifications to the system that can be corroborated by experiment.\cite{he} In this term paper we take a brief look at applying such methods to understand the origin of the observed ionic conductivity in $c-LLZO$ and modifications to the structure that may help us optimize its properties.

\section*{\normalfont Structure of Cubic $Li_7La_3Zr_2O_{12}$}

\begin{figure}[t]
	\centering
	\begin{subfigure}[b]{0.3\textwidth}
		\centering
		\includegraphics[height=2.4in]{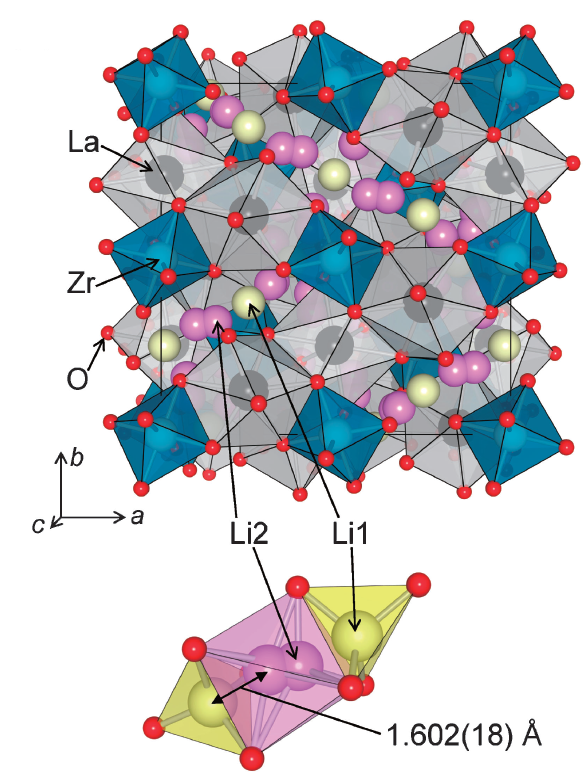}
		\caption{{}}
	\end{subfigure}
	\begin{subfigure}[b]{0.3\textwidth}
		\centering
		\includegraphics[width=2in]{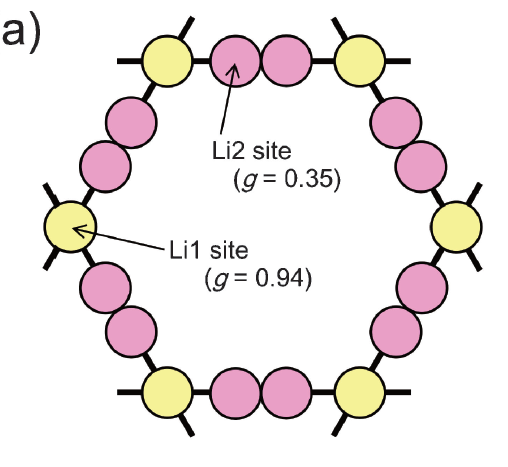}
		\caption{{}}
	\end{subfigure}
	\begin{subfigure}[b]{0.3\textwidth}
		\centering
		\includegraphics[width=2in]{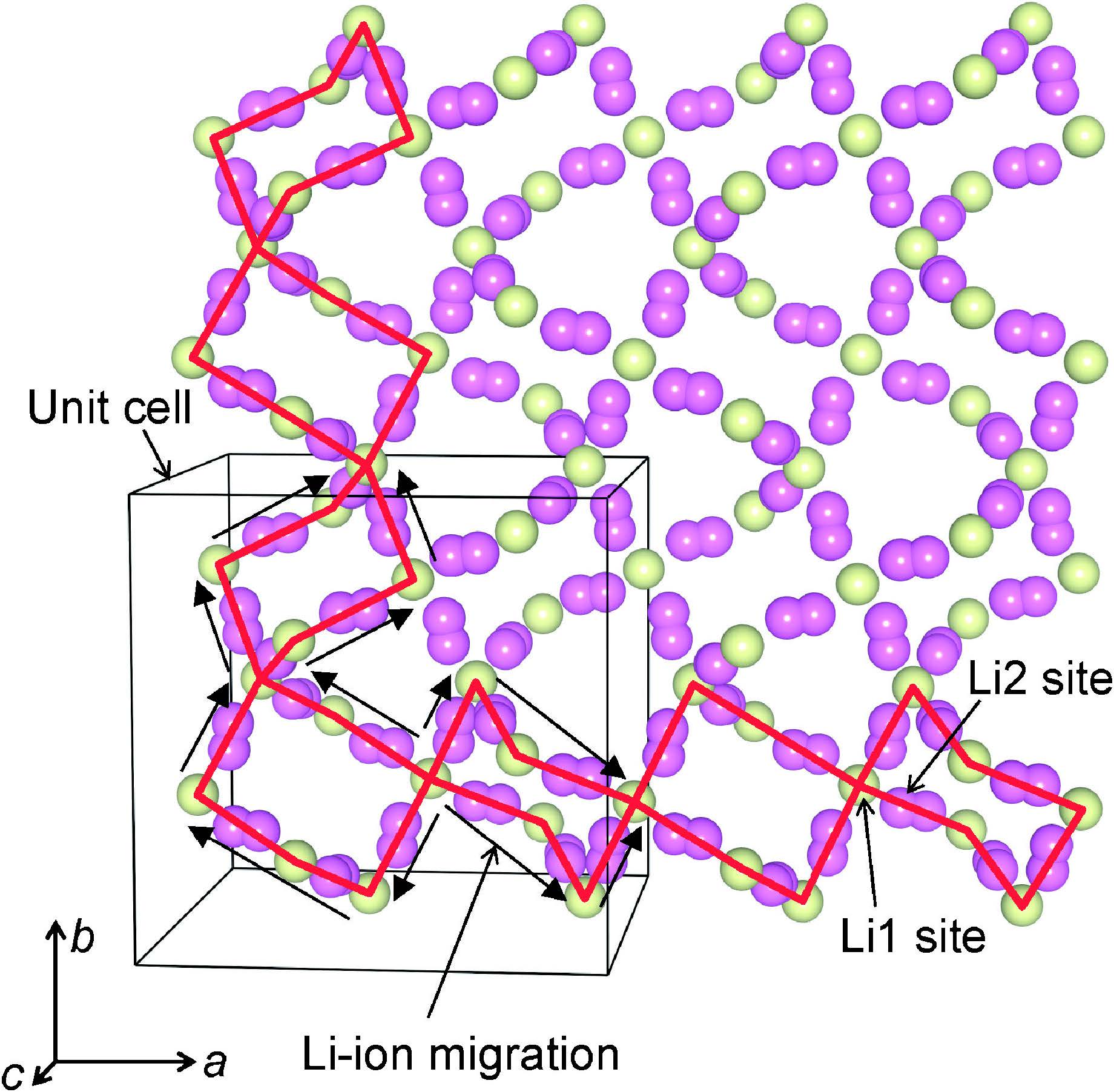}
		\caption{{}}
	\end{subfigure}
\caption{\textbf{\textit{(a)}} The $c-LLZO$ unit cell and showing the coordination polyhedra around the $Li1$ and $Li2$ sites.\cite{awaka}
\textbf{\textit{(b)}} The $Li^+$ loop structures constructed in $c-LLZO$ with the occupancy values $(g)$ indicated.\cite{awaka}
\textbf{\textit{(c)}} $3D$ network of lithium conduction pathways proposed by Awaka et al.\cite{awaka}
} 
\end{figure}

$LLZO$ exhibits two stable polymorphs, a cubic and tetragonal phase.\cite{murugan}\cite{bbuschmann} The cubic phase exhibits a higher ionic conductivity compared to the tetragonal phase, therefore is of greater interest.\cite{wang}\cite{murugan} Throughout this discussion, unless explicitly specified we will focus our attention on the cubic phase. The cubic phase has a space group $Ia\bar{3}d$ and a lattice constant $a=12.9827\textnormal{\AA}$ but can vary anywhere from $\sim12-13\textnormal{\AA}$ depending on doping and its concentration.\cite{murugan}\cite{jalem} Neutron diffraction studies have confirmed that the $Li^+$ ions occupy three distinct kinds of lattice sites, 24 tetrahedral $(24d)$, 48 octahedral $(48g)$ and 96 off-centre octahedral $(96h)$.\cite{wang}\cite{murugan}\cite{awaka} These sites are face-shared and form a $3D$ network.\cite{awaka} The cubic phase also exhibits a disordered $Li^+$ ion distribution with 56 $Li^+$ ions in each unit cell and presence of empty tetrahedral and octahedral sites.\cite{awaka} Figure 1 shows a $c-LLZO$ unit cell showing the different possible $Li^+$ ion occupation sites.\cite{awaka}  This reported disordering and the presence of unoccupied lithium sites play a key role in conduction.\cite{he}\cite{awaka}\cite{jalem} It has also been observed that some octahedral sites have a lower occupancy compared to others due to close proximity to each other $1.6\textnormal{\AA}$ causing coulombic repulsion.\cite{wang}\cite{he}\cite{awaka} The $Li^+$ ions can occupy principally 48 of the 96 distorted octahedral sites.\cite{he}\cite{awaka} Occupancy of the total population of ions is distributed roughly tetrahedral $(40\%)$ and octahedral $(60\%)$ in their lowest potential energy configuration, showing a slight preference for the octahedral $(48g/96h)$ sites.\cite{he}

\subsection*{\normalfont Migration Mechanism \& Kinetics}

\begin{figure}[t]
	\centering
	\begin{subfigure}[b]{1\textwidth}
		\centering
		\includegraphics[height=3.4in]{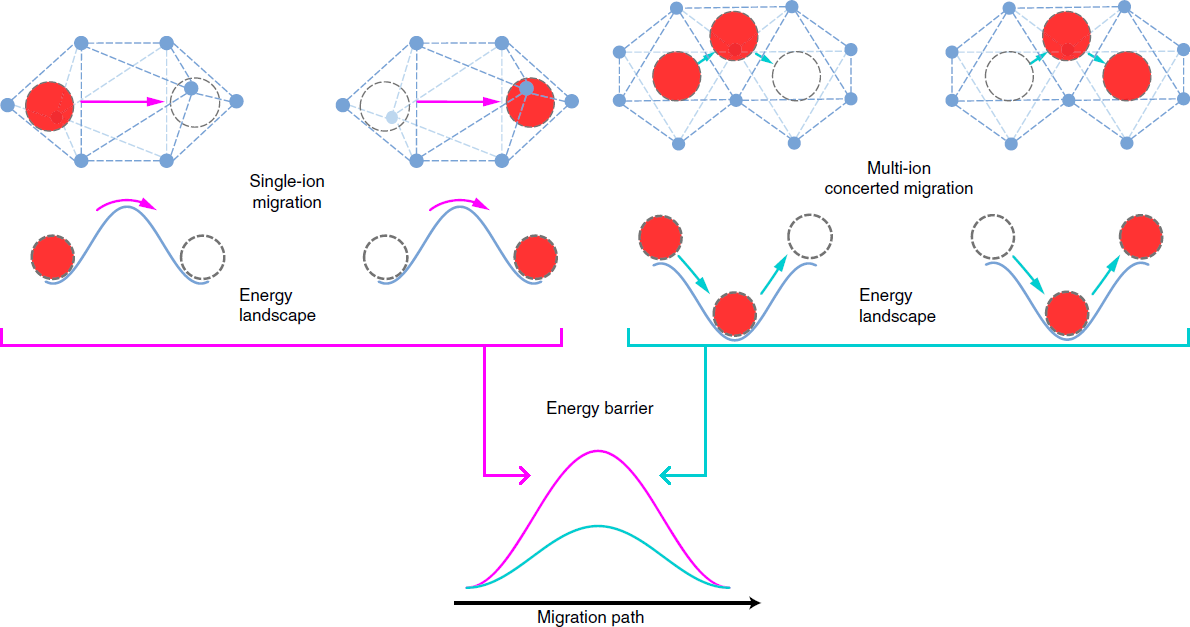}
		\caption{{}}
	\end{subfigure}
\caption{\textbf{\textit{(a)}} Schematic representation of the energy barrier profiles in the case of single-ion migration and multi-ion concerted migration.\cite{he}  }

\end{figure}

\begin{figure}[t]
	\centering
	\begin{subfigure}[b]{0.3\textwidth}
		\centering
		\includegraphics[height=3.4in]{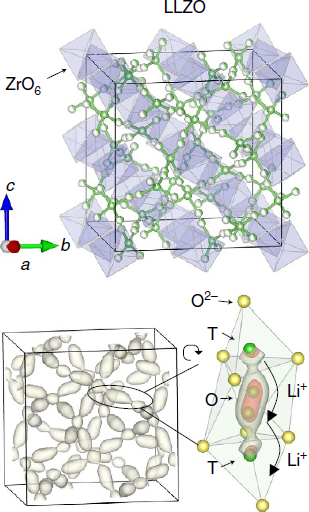}
		\caption{{}}
	\end{subfigure}
	\begin{subfigure}[b]{0.3\textwidth}
		\centering
		\includegraphics[height=3.4in]{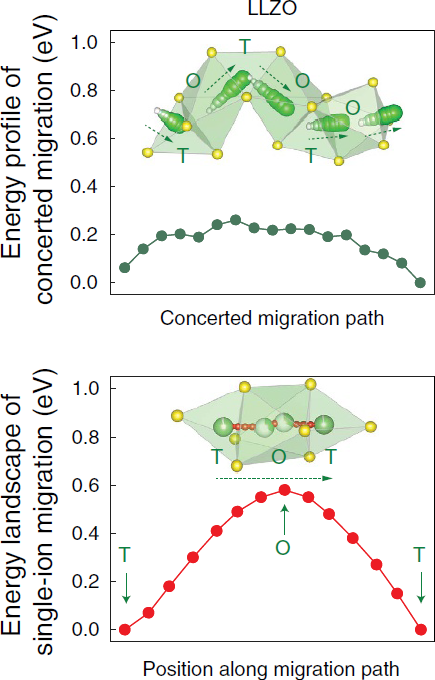}
		\caption{{}}
	\end{subfigure}
	\begin{subfigure}[b]{0.3\textwidth}
		\centering
		\includegraphics[height=3.4in]{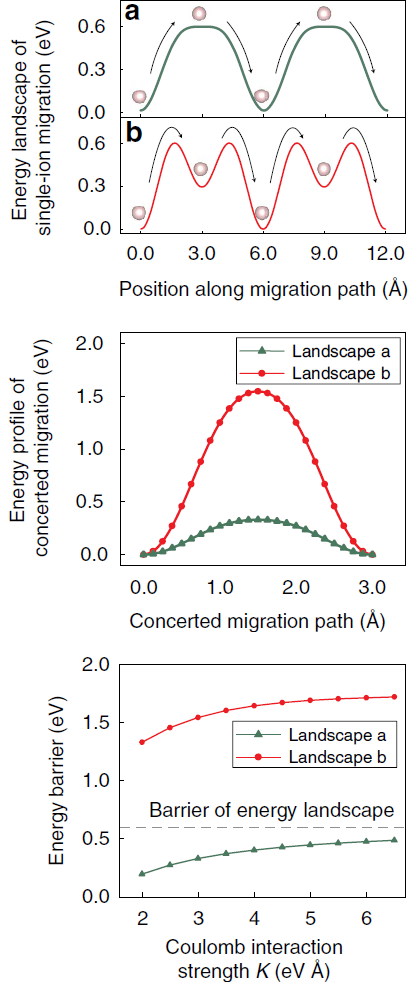}
		\caption{{}}
	\end{subfigure}

\caption{\textbf{\textit{(a)}} The structure of $c-LLZO$ and the spatial energy profile of the migration pathways in the unit cell as mapped out by the occupancy probability of the mobile $Li^+$ ions in the $AIMD$ simulation from high energy sites to low energy sites.\cite{he} The $Li^+$ ions in the unit section $(T-O-T)$ comprising the migration pathway are shown in detail.\cite{he}
\textbf{\textit{(b)}} A schematic representation of the migration process and the calculated energy barrier profile using $NEB$ for both concerted multi-ion migration and single-ion migration.\cite{he} One can clearly observe a flatter energy profile for multi-ion concerted migration.\cite{he}
\textbf{\textit{(c)}} The potential energy of the structural framework at the high energy sites.\cite{he} The energy profile of the multi-ion migration calculated for profile \textbf{a} and \textbf{b}.\cite{he} The barrier as a function of the coulomb interaction strength $K$ for landscape \textbf{a} and \textbf{b}.\cite{he}}
\end{figure}

The migration of $Li^+$ ions is the primary phenomenon responsible for the observed high conductivity of $c-LLZO$ compared to other solid state electrolytes.\cite{murugan}\cite{meier}\cite{he} However, the actual mechanism of migration is still a matter of great debate amongst researchers.\cite{meier}\cite{he} This interest in unravelling this mechanism was sparked by the determination of the crystal structure of $c-LLZO$ by Awaka et al, who proposed a possible migration pathway for the $Li^+$ ions as can be seen in Figure 1.\cite{awaka} Since then numerous studies have been conducted on mapping the energy landscape and feasibility of various ion migration pathways in such a structure.\cite{wang}\cite{meier}\cite{he}\cite{jalem}

He et al, used \textit{ab-initio} molecular dynamics $(AIMD)$ simulations to study the energy landscape of possible $Li^+$ ion migration pathways in solid state electrolytes, primarily $c-LLZO$.\cite{he} The proposed mechanism that he simulates is the concerted migration of $Li^+$ ions in the structure through pathways composed of face-connected tetrahedral and octahedral sites as can be seen in Figure 3.\cite{he} The $NEB$ simulated activation barrier for such a mechanism $\sim0.26\;eV$ is also in much better agreement with experimental results $\sim0.3\;eV$, compared to the $0.58\;eV$ diffusion barrier predicted for single ion hops as seen in Figure 3.\cite{he} The disordered lithium ion distribution also stipulates the existence of multiple ion migration pathways and $Li^+$ ion configurations, this was accounted for by performing multiple $AIMD$ simulations with different initial ion configurations which are also in close agreement with these results.\cite{he} Therefore, local environment of the migrating ions has an effect on the activation barrier, however these effects are not large enough to change the migration mechanism underneath.\cite{he} This analysis also reveals that a classical diffusion based model, as seen in Figure 3 and 4, is ineffective in explaining the observed low activation barrier, and that it is the collective effect of strong mobile-ion interactions; primarily coulombic; combined with the partial occupation of high energy sites; primarily octahedral $(48g)$ and distorted octahedral $(96h)$ sites by excess $Li^+$ ions in stuffed garnets, which results in multi-ion diffusion being the favoured mechanism.\cite{wang}\cite{he} High energy sites also have locally flatter energy landscapes due to the fact that during concerted migration high energy ions migrate downhill which offsets the energy cost of the ions migrating uphill.\cite{wang}\cite{he} This facilitates easier movement of $Li^+$ ions, as is observed in the $AIMD$ simulations.\cite{he} This results in an elongated spatial occupation density of $Li^+$ ions, as can be seen in Figure 3, allowing for easier $Li^+$ ion hopping and facilitating multi-ion migration.\cite{he}
This mechanism provides us a theoretical basis for designing better $Li^+$ ion conductors.\cite{he} By inserting mobile $Li^+$ ions into high energy sites, the energy landscape of the migration pathways can be manipulated; allowing the ions to migrate more freely.\cite{wang}\cite{he} This will facilitate concerted multi-ion migration over single-ion diffusion mediated conduction which is energetically unfavourable.\cite{he} Another strategy that can be employed to manipulate the energy landscape of the electrolyte is addition of dopant atoms which are usually heavier, less mobile and greatly modify the energy landscape of the migrating ions.\cite{he}Such a framework will allow the targeted optimization allowing us to design electrolytes with higher conductivities at room temperature.\cite{wang}

\section*{\normalfont Conclusion}
In this term paper we looked at $AIMD$ studies that propose a possible model for the underlying mechanism of $Li^+$ ion migration in $c-LLZO$. From the presented results and their agreement with experimental evidence it is reasonable to conclude that the primary cause for the observed high ionic conductivity in $c-LLZO$ is the concerted multi-ion migration of lithium through pathways stabilized by low activation barriers, mobile-ion coulombic interactions and ion-lattice coulombic interactions.\cite{wang}\cite{he} $NEB$ studies of the multiple possible $Li^+$ ion configurations also point to a significantly lowered migration activation barrier, the cause of which is pointed to the downhill migration $(O\rightarrow T)$ of the $Li^+$ ions in high energy lattice sites compensating for the required activation energy for uphill migration $(T \rightarrow O)$ of $Li^+$ ions at low energy lattice sites.\cite{he} This results in an overall flatter energy landscape, than would be predicted by single-ion diffusion mediated migration.\cite{he} This makes it easier for excess lithium ions in the lattice to move through the migration pathways, explaining the comparatively high experimental ionic conductivity.\cite{he} Such a model provides us with the theoretical basis for the observed phenomenon and a targeted approach to optimizing the behaviour of these materials.\cite{wang}\cite{he} This will prove to be an indispensable tool for further study on the various possible approaches to making this technology a commercial reality.\cite{wang}
\bibliography{myref}

\begin{thebibliography}{1}

\bibitem{wang}
Chengwei Wang, Kun Fu, Sanoop~Palakkathodi Kammampata, Dennis~W. McOwen,
  Alfred~Junio Samson, Lei Zhang, Gregory~T. Hitz, Adelaide~M. Nolan, Eric~D.
  Wachsman, Yifei Mo, Venkataraman Thangadurai, and Liangbing Hu.
\newblock {Garnet-Type Solid-State Electrolytes: Materials, Interfaces, and
  Batteries}.
\newblock {\em Chem. Rev.}, 120(10):4257--4300, May 2020.

\bibitem{murugan}
Ramaswamy Murugan, Venkataraman Thangadurai, and Werner Weppner.
\newblock {Fast Lithium Ion Conduction in Garnet-Type
  $Li_{7}La_{3}Zr_{2}O_{12}$}.
\newblock {\em Angewandte Chemie International Edition}, 46(41):7778--7781,
  October 2007.

\bibitem{cussen}
Edmund~J. Cussen.
\newblock {Structure and ionic conductivity in lithium garnets}.
\newblock {\em J. Mater. Chem.}, 20(25):5167--5173, 2010.

\bibitem{xie}
Hui Xie, Jose~A. Alonso, Yutao Li, Maria~T. Fernández-Díaz, and John~B.
  Goodenough.
\newblock {Lithium Distribution in Aluminum-Free Cubic
  $Li_{7}La_{3}Zr_{2}O_{12}$}.
\newblock {\em Chem. Mater.}, 23(16):3587--3589, August 2011.

\bibitem{meier}
Katharina Meier, Teodoro Laino, and Alessandro Curioni.
\newblock {Solid-State Electrolytes: Revealing the Mechanisms of Li-Ion
  Conduction in Tetragonal and Cubic LLZO by First-Principles Calculations}.
\newblock {\em J. Phys. Chem. C}, 118(13):6668--6679, April 2014.

\bibitem{he}
Xingfeng He, Yizhou Zhu, and Yifei Mo.
\newblock {Origin of fast ion diffusion in super-ionic conductors}.
\newblock {\em Nature Communications}, 8(1):15893, 2017.

\bibitem{awaka}
Junji Awaka, Akira Takashima, Kunimitsu Kataoka, Norihito Kijima, Yasushi
  Idemoto, and Junji Akimoto.
\newblock {Crystal Structure of Fast Lithium-ion-conducting Cubic
  $Li_{7}La_{3}Zr_{2}O_{12}$}.
\newblock {\em Chem. Lett.}, 40(1):60--62, November 2010.

\bibitem{bbuschmann}
Henrik Buschmann, Janis Dölle, Stefan Berendts, Alexander Kuhn, Patrick
  Bottke, Martin Wilkening, Paul Heitjans, Anatoliy Senyshyn, Helmut Ehrenberg,
  Andriy Lotnyk, Viola Duppel, Lorenz Kienle, and Jürgen Janek.
\newblock {Structure and dynamics of the fast lithium ion conductor
  “$Li_{7}La_{3}Zr_{2}O_{12}$”}.
\newblock {\em Phys. Chem. Chem. Phys.}, 13(43):19378--19392, 2011.

\bibitem{jalem}
Randy Jalem, Yoshihiro Yamamoto, Hiromasa Shiiba, Masanobu Nakayama, Hirokazu
  Munakata, Toshihiro Kasuga, and Kiyoshi Kanamura.
\newblock {Concerted Migration Mechanism in the $Li$ Ion Dynamics of
  Garnet-Type $Li_{7}La_{3}Zr_{2}O_{12}$}.
\newblock {\em Chem. Mater.}, 25(3):425--430, February 2013.

\end{thebibliography}
\bibliographystyle{unsrt}

\end{document}